\documentstyle[12pt]{article}
\def\ba{\begin{array}}
\def\ea{\end{array}}
\def\be{\begin{equation}\begin{array}{l}}
\def\ee{\end{array}\end{equation}}
\def\bea{\begin{equation}\begin{array}{l}}
\def\eea{\end{array}\end{equation}}
\def\f{\frac}

\def\de{\delta}

\def\si{\Sigma}
\def\t{T_+}
\def\v{v_{cl}}
\def\m{m_{cl}}
\def\n{N_{cl}}
\def\bi{\bibitem}
\def\c{\cite}
 
\title{ {\bf Evaporation of Schwarzschild Black Hole in the Large N Matrix Theory} }
\author{ Yi-Hong Gao and Wei Zhang \thanks{zhwei@itp.ac.cn} \\
 Institute of Theoretical Physics , Academia Sinica,\\
 P.O.Box 2735, Beijing, 100080, China  }
\date{ }
\begin{document}
\maketitle

\vskip 0.8in

\begin{center}{\large \bf Abstract} \end{center}
\quad
By using the D0-brane cluster picture, we consider the Hawking radiation of Schwarzschild black hole (SBH) in large N Matrix model. We get the correct formula for the the Hawing evaporation rate. Our results give some evidence on the Lorentz invariance of the physics of Matrix model.

%\end{center}
\newpage

\baselineskip 18pt
\section{Introduction}
\quad
Recently, much attention has been paid on the Matrix model \c{bf} \c{se} description of the basic qualitative features of Schwarzschild black hole (SBH) \c{ba1} \c{ba2} \c{kl} \c{ml} \c{ml2} \c{ho} \c{ml3} \c{mi1} \c{ram} \c{gz} \c{low}. The basic idea of Banks, Fischler, Klebanov and Susskind (BFKS) is to describe SBH as Boltzmann D0-brane gas \c{ba1}\c{ba2}. Under the assumption of $N \sim S$, they were able to reproduce the Bekenstein-Hawking entropy of the black hole by using some knowledge of thermodynamics of supersymmetric Yang-Mills field (in D=8). Banks, Fischler, Klebanov (BFK) \c{ba3} calculated the Hawking evaporation rate correctly. Their result in fact relies on the Lorentz invariance of Matrix model which deserves further investigation. For the case $N>>S$, the Bekenstein-Hawking entropy was derived in \c{ml}\c{gz}\c{low}. In this paper, we will consider the Hawking radiation in Matrix model for the case $N>>S$. We are able to get the correct Hawking evaporation rate. The covariance of our result gives some evidence on the Lorentz invariance of the physics of Matrix model.

We will adopt the technique of BFK. The main point of our approach is that D0-brane bound state will emit cluster (correlated domain) of D0-branes, which was proposed by Horowitz, Martinec and Li \c{ml2}\c{ho}. As explained in \c{ho}, the transverse size of a black hole remains constant under longitudinal boost. According to weak holographic principle, the degree of freedom of the theory is proportional to the area of the (D-2)-dimensional spatial surface. When $N>>S$, clusters  of D0-branes will form. The effective degree of freedom will not increase any more. In section 2 we will give a brief review of the cluster picture of D0-brane gas in large N matrix model. Section 3 and section 4 are about Hawing radiation in cluster picture. \\

\section{Cluster picture of D0-brane gas in large N Matrix model}  
\quad
In the cluster picture, the number of cluster is $\n=S$. Each cluster is composed of $\f{N}{S}$ D0-branes. The mass of the cluster is $\f{N}{SR}$ (R is the radius of the longitudinal dimension) and the transverse velocity of cluster is $\f{SR}{NR_s}$ ($R_s$ is the radius of SBH)\c{ml}. Thus the momentum of the cluster is $\f{1}{R_s}$, which is consistent with the uncertainty principle. The temperature of the cluster gas is $\t=\f{E}{S}$. With the help of energy relationship $E=\f{M^2R}{N}$ and $S=MR_s$, 
\bea
\t=\f{MR}{NR_s}\\
=\f{1}{2}\m \v^2
\eea  

The interaction among the clusters is of the form
\be
 \si_{a,b=1}^S G_D \f{(\m\v^2)_a (\m\v^2)_b}{R r^{D-4}}
\ee
Taking into account the interaction in which the clusters exchange longitudinal momentum, one can get the correct thermodynamics of SBH in the cluster picture of large N Matrix model \c{ml2}.   

As argued in \c{ba1}, the typical longitudinal momentum of Hawking particle is
\bea
p_-=\f{N}{MR} \f{1}{R_s}\\
=\f{N}{S} \f{1}{R}
\eea
This means the Hawking particle is composed of $\f{N}{S}$ D0-branes. This is nothing but that the Hawking particle is just "cluster" of D0-branes. Consider a SBH with radius $R_s$, which is described in Matrix theory by a cluster bound state in transverse spatial dimension. It is those clusters that are within the crust of depth $\v \de t$ can escape from the cluster bound state within the time $\de t$. So the probability per unit time to emit a cluster is $\f{\v}{R_s} \sim \t$. Since each cluster has energy $\f{1}{2}\m \v^2 \sim \t$, the rate of energy loss in light-cone time is $\t^2$.
More carefully treatment will be given in the following sections.
 
\section{Calculation of the Hawking evaporation rate}
\quad
The probability per unit time to emit a cluster is given by \c{ba3}
\be
\f{d\n}{dx^+} \sim S \f{N}{SR} \si_{n>0} \int_0^{\infty} dp_+ \int d^{D-2}p_- \de (n\f{
N}{SR}-\f{p^2}{p_+}) \times |A(n,y)|^2
\ee
where A(n,y) is the wavefunction of a single cluster, p is the transverse momentum. According to the argument of BFK, 
\be
|A(n,y)|^2 \sim G_D
\ee
and the n is of order 1. Performing the integration leads to
\be
\f{d\n}{dx^+} \sim G_D \f{RS^2}{N} p^D
\ee
Using the fact that the typical momentum in transverse dimension is
\be
p \sim \f{1}{R_s}
\ee
and
\be
S=\f{R^{D-2}_s}{G_D}
\ee
we have
\be
\f{d\n}{dx^+} \sim \t
\ee
The conventional semiclassical calculation in light-cone frame is almost the same as that done by \c{ba3}.
\be
\f{d\n}{dx^+} \sim \int_0^{\infty} dp_- \inf_0^{\infty} dp_+ \int d^{D-2}p \de (p_+p_--P^2) \times e^{-p_+/T_+}e^{-p_-/T_-} A p_+
\ee
where $A=4G_DS$ is the horizon area. The only change is the value for $\t, T_-$. In our case, the boost factor is $\f{MR}{N}$. The temperature $\t ,T_-$ is
\bea
\t=\f{MR}{NR_s}\\
T_-=\f{N}{MRR_s}
\eea
The result of semiclassical calculation is the same as (9).

\section{Calculation of the rate of mass loss}
\quad
The formula for the rate of light cone energy loss per unit light cone time is
\be
\f{dE}{dx^+} \sim S \f{N}{SR} \si_{n>0} \int_0^{\infty} dp_+ \int d^{D-2}p_- \de (n\f{
N}{SR}-\f{p^2}{p_+}) \times p_+ |A(n,y)|^2
\ee
A bit calculation leads to
\be
\f{dE}{dx^+} \sim \t^2
\ee
From the relation
\be
E=\f{M^2R}{N}
\ee
we have
\be
\f{dM}{dx^+} \sim \f{E}{R}\f{dN}{dx^+} +\f{N}{R}\f{dE}{dx^+}
\ee
Using the fact 
\be
\f{dN}{dx^+}=\f{N}{S}\f{d\n}{dx^+}
\ee
(13) can be written as
\be
\f{dM}{dx^+}=\f{1}{R_s} \t
\ee
Boosting back to the rest frame, the rate of mass loss per unit time is 
\be
\f{dM}{dx^+_{rest}}=\f{1}{R^2_s}
\ee

Above calculations show that our result about Hawking radiation is independent of the boost parameter. Thus our result in some sense gives some support of the Lorentz invariance of Matrix model

%{ \large \bf Acknowledgment}\\
\newpage
%\begin{flushleft}
%\baselineskip 24pt
%{\large \bf References}\\

%\end{flushleft}
\end{document}